\documentclass[%
 aip,
 jmp,%
 amsmath,amssymb,amsfonts,amsthm,
 reprint,%
]{revtex4-1}
\usepackage{slashed}
\usepackage{chngcntr}

\counterwithin*{equation}{section}
\counterwithin*{equation}{section}
\usepackage{graphicx}
\usepackage{dcolumn}
\usepackage{bm}

\begin{document}


\title[Noncommutative Spaces and the Theory of Geometrodynamics]{Noncommutative Spaces and the Theory of Geometrodynamics}

\author{Nikhil Kalyanapuram}%
 \email{nikhilkaly@cmi.ac.in}
\affiliation{ 
Chennai Mathematical Institute, H1, SIPCOT, Siruseri, Kelambakkam, Tamil Nadu. \\
Centre for Fundamental Research and Creative Education, Bangalore, Karnataka.
}%

\date{\today}

\begin{abstract}
This article is concerned with a generalisation of Connes' noncommutative framework. This is achieved by a general study of spectral triples, in particular through an analysis of the role played by the Dirac operator. The Dirac operator is used to construct a generalised covariant derivative, which is then employed to define a generalised spacetime. The spectral action is then modified to accommodate this generalisation and it is observed that in the appropriate limit, Connes' spectral action is recovered.
\end{abstract}

\maketitle

\begin{quotation}

\end{quotation}

\section{Introduction}
The application of geometry to describe physical phenomena and thus achieve a geometrical foundation for theoretical physics is a means to ensure a purely axiomatic formulation of the mathematical descriptions of the phenomena seen in nature. Geometry, being of a very intuitive nature, affords to us a means to extend our intuitions of the world to a more solid foundation built upon mathematics. 

The earliest development of this kind is that of Euclid and his geometry of flat spaces. The Euclidean geometrical apparatus provides a firm grounding for many of the principles of classical mechanics. Basic notions of distance and displacement, for example, can be formulated using the familiar Euclidean distance formula, whose two-dimensional version coincides with the Pythagorean theorem. The vector analysis upon which classical mechanics rests can be formulated on a three dimensional Euclidean space $\mathbb{R}^{3}$.  

Formulating the theory of special relativity\cite{ein3} in a mathematically rigorous fashion has been made feasible after Minkowski indroduced his notion of a four-dimensional manifold, now called Minkowski space, wherein the a formal equivalence between space and time co-ordinates is achieved. Even so, this space does not have any curvature, and the distances between any two points are still calculated, through the generalised Euclidean formula.

The general principle of relativity, the idea that physical laws are invariant under any arbitrary transformation of co-ordinates is given validity only under the framework offered by Riemannian geometry, a geometry which allows the manifold to possess a curvature. This curvature of space however, if interpreted as the origin of gravity, offers a purely geometric view of this phenomenon, and this proposition forms the basis of general relativity\cite{ein1}.

In the following paper, we attempt to construct a geometrical theory using the non-commutative geometry developed by Alain Connes\cite{connes2} to  extend Riemannian geometry into a more general metric framework and use this system to achieve a geometric interpretation of gravity, gauge interactions, the Higgs field and matter fields.

\section{The Geometry of General Relativity}
A Riemannian manifold, described by an ordered pair $\left(M. g\right)$, where $M$ is a smooth manifold equipped with a Riemannian metric $g$ which satisfies properties of positive definiteness and non-degeneracy. The metric tensor, described by a symmetric matrix, is used to define the infinitesimal distance using the following formula,

\begin{equation}
\mathrm{d}s^{2} = g_{\mu\nu}\mathrm{d}x^{\mu}\mathrm{d}x^{\nu} 
\end{equation}

The inverse of the above metric tensor $g_{\mu\nu}$ is the contravariant metric $g^{\mu\nu}$. 

The Einstein-Hilbert action, the action for general relativity is then defined as,

\begin{equation}
S_{EH} = \int \left(\frac{1}{2\kappa}R + \mathcal{L}_{M}\right) \sqrt{-g}\mathrm{d}^{n}x
\end{equation}

In the above Lagrangian, gravitational dynamics are assumed to be encapsulated in the Ricci scalar. Here $ \mathcal{L}_{M}$ is the collective Lagrangian of all matter and energy fields, $\kappa$ is a constant and $g$ is the determinant of the metric. Taking the variation with respect to $g^{\mu\nu}$, we obtain the Einstein field equations as,

\begin{equation}
R_{\mu\nu} - \frac{1}{2}g_{\mu\nu}R = \kappa T_{\mu\nu}
\end{equation}

where $T_{\mu\nu}$, the stress-energy tensor is given by,

$$
T_{\mu\nu} = -2\frac{\delta  \mathcal{L}_{M} \sqrt{-g}}{\delta g^{\mu\nu}}
$$

The Einstein field equation allows us to envision gravity as the geometric curvature of space and time, caused by the presence of matter and energy. By making a change to the Einstein field equations, of the form,

\begin{equation}
S_{EH} = \int \left(\frac{1}{2\kappa}R + \mathcal{L}_{M} + 2\gamma_{0}\right) \sqrt{-g}\mathrm{d}^{n}x
\end{equation}

where $\gamma_{0}$ is the vacuum energy, a constant value, the Einstein field equations obtain a term contributing to the curvature of space on account of the energy of the vacuum.

General relativity allows us to make co-ordinates superfluous inasmuch as their physical interpretations are concerned.

\section{Noncommutative Geometry}

\subsection{Spectral Triples}
The basic information and data regarding a non-commutative space, as has been developed by Alain Connes have been enshrined in what is known as a spectral triple. A spectral triple is an ordered triplet $\left(\mathcal{A}, D, \mathcal{H}\right)$. $\mathcal{A}$ is an algebra of operators on the Hilbert space $\mathcal{H} = L^{2}\left(M, S\right)\otimes\mathcal{H}_{F}$ of $L^{2}$-spinors, $D$ is a Dirac operator acting on $\mathcal{H}$ such that $D$ has a compact resolvent and the commutators $\left [ D,a \right ]$ are bounded for all $a\in\mathcal{A}$. $\mathcal{H}_{F}$ is a Hilbert space with a basis labelled by elementary fermions. 

A spectral triple has a Dirac operator $D$ which in the commutative case, which will be elaborated upon later, is of the form,

\begin{widetext}
\begin{equation}
D = \frac{1}{\sqrt{-1}}\gamma^{\mu}\left(\partial_{\mu} - \frac{1}{4}\omega_{\mu}\right) \otimes \mathbb{I} + \gamma^{5}\otimes D_{Y} = \slashed{\partial}_{M}\otimes \mathbb{I} + \gamma^{5}\otimes D_{Y}
\end{equation}
\end{widetext}

where $\omega_{\mu}$ is the spin connection and $D_{Y}$ is given by,

$$
D_{Y} = \begin{bmatrix}
Y & 0\\ 
0 & \overline{Y}
\end{bmatrix}
$$

where $Y$ is a Yukawa coupling matrix.

A Riemannian metric is defined on the space through the following heuristic definition,

\begin{equation}
\mathrm{d}s^{2} = D^{-2} = \mathrm{d}x^{\mu}g_{\mu\nu}\left(\mathrm{d}x^{\nu}\right)^{*}
\end{equation}

$$
D = \frac{1}{\sqrt{-1}}\gamma^{\mu}\nabla_{\mu}
$$

where $g_{\mu\nu}$ is a positive definite matrix. The above definition of the line element is important for two reasons. The first reason is that it heuristically determines the metric used to define distances on the space. The second has to do with the quantum nature of fermions. According to the identity for Grassmannian spinors $\psi$,

\begin{equation}
\int \mathcal{D}\psi\mathcal{D}\overline{\psi}\left \{ i\int\overline{\psi}D\psi \mathrm{d}^{n}x \right \} = D^{-1}
\end{equation}

we see that the propagator for the fermions of the space acts as the line element. This draws a relation between the spinors and spacetime.

Apart from the earlier definition of the spectral triple, there is an addendum to the definition that must be elaborated upon. If the Hilbert space is endowed with a $\mathbb{Z}$ modulo 2 grading $\gamma$ having the following properties,

$$
\gamma = \gamma^{*}
$$

$$
\gamma^{2} = 1
$$

$$
\gamma a = a \gamma \;\;\; \forall \mathcal{A}
$$

$$
\gamma D = - D \gamma
$$

the spectral triple is known as an {\it even} spectral triple.

The non-commutative space is endowed with an antilinear isometry $J: \mathcal{H}\rightarrow\mathcal{H}$ such that $J$ satisfies the following properties,

$$
J^{2} = \epsilon
$$

$$
JD = \epsilon^{'}DJ
$$

$$
J\gamma = \epsilon^{''}\gamma J
$$
where $\epsilon, \epsilon^{'}, \epsilon{''} \in \left \{ -1,1 \right \}$. 

\subsection{Inner Fluctuations and The Dirac Operator}
If the algebra of operators on the Hilbert space is taken to be non-commutative, it admits natural internal fluctuations which give rise to the gauge bosons and the Higgs field of the Standard Model when appropriate choices for $\mathcal{A}$ and $D_{F}$ are made. For the Standard Model, we have for $\mathcal{A}$,

\begin{equation}
\mathcal{A} = C^{\infty}\left(M\right)\otimes \mathcal{A}_{F}
\end{equation}

$$
\mathcal{A}_{F} = \mathbb{C}\oplus\mathbb{H}\oplus M_{3}\left(\mathbb{C}\right)
$$

where $C^{\infty}\left(M\right)$ is the algebra of smooth functions on $M$, $\mathbb{C}$ is the set of complex numbers, $M_{3}\left(\mathbb{C}\right)$ is the set of $3\times 3$ matrices on $\mathbb{C}$ and $\mathbb{H}$ is the algebra of quaternions,

$$
\mathbb{H} = \left \{ \begin{pmatrix}
x & y\\ 
-y^{*} & x^{*}
\end{pmatrix} : x,y\in\mathbb{C} \right \}
$$

The inner fluctuations of the algebra $\mathcal{A}$ parametrise the gauge bosons and the Higgs field as we shall soon see. The inner fluctuations $A$ are elements of the set $\Omega^{1}_{D}$,

\begin{equation}
\Omega^{1}_{D} = \left \{ \sum a_{i}\left [ D, b_{i} \right ] : a_{i}, b_{i}\in\mathcal{A}\right \}
\end{equation}

The Dirac operator, currently possessing only terms that allow the fermions to be coupled to gravity, must be perturbed in order to allow for coupling with the gauge fields as well. The Dirac operator is thus perturbed according to the formula,

\begin{equation}
D^{'} = D + A + JAJ^{-1}
\end{equation}

Since the Dirac operator splits into a sum of two parts $\slashed{\partial}_{M}$ and $D_{Y}$, $A$ also splits into a discrete part $A^{0,1}$ and a continuous $A^{1,0}$. The discrete part is obtained by the computation 

$$
\sum a_{i}\left [\gamma^{5}\otimes D_{F}, b_{i}\right]
$$

and the continuous part is obtained by the computation,

$$
\sum a_{i}\left [\slashed{\partial}_{M}, b_{i}\right]
$$

The Yukawa operator $D_{Y}$ is given by,

$$
D_{Y} = \begin{bmatrix}
Y & 0\\ 
0 & \overline{Y}
\end{bmatrix}
$$

with,

$$
Y = Y_{q}\otimes \mathbb{I}_{3} + Y_{\ell}
$$

$$
Y_{q} = \begin{bmatrix}
0 & k^{d} & i\sigma_{2}k^{u}\\ 
\left ( k^{d} \right )^{*} & 0 & 0\\ 
i\sigma_{2}\left ( k^{u} \right )^{*} & 0 & 0
\end{bmatrix} 
$$

$$
Y_{\ell} = \begin{bmatrix}
0 & k^{e} \\ 
i\sigma_{2}\left ( k^{e} \right )^{*} & 0 \ 
\end{bmatrix} 
$$

where $k^{u}$, $k^{d}$ and $k^{e}$ are Yukawa matrices for the up quark family, down quark family and electron family respectively, with the generations contained in the operators itself.

Under the algebra chosen, and choosing $a = \left(\lambda, q, m\right)$ and $b=\left(\lambda^{'}, q^{'}, m^{'}\right)$, we obtain an $U(1)$ field $\Lambda$, an $SU(2)$ field $Q$ and an $U(3)$ field $V$ according to,

\begin{equation}
\Lambda = \sum \lambda\left [\slashed{\partial}_{M}, \lambda^{'}\right]
\end{equation}

\begin{equation}
Q = \sum q\left [\slashed{\partial}_{M}, q^{'}\right]
\end{equation}

\begin{equation}
V = \sum m\left [\slashed{\partial}_{M}, m^{'}\right]
\end{equation}

A gauge potential $A$ is said to be unimodular iff $\mathrm{Tr}(A)$ vanishes. In our case, by imposing the unimodularity condition, $Q$ identically vanishes, giving us,

\begin{equation}
\mathrm{Tr}(V) + \Lambda = 0
\end{equation}

This in turn gives us,

$$
\mathrm{Tr}(V) = -\Lambda 
$$

which we can write as a sum,

\begin{equation}
V = -V^{'} - \frac{1}{3}\Lambda \mathbb{I}_{3}
\end{equation}

where $V^{'}$ is an $SU(3)$ gauge potential.

The discrete fluctuations $A^{0,1}$ are however parametrised by quaternion valued functions $H$ and $H^{*}$, the computation of which is given in Section 3.5.1 and Proposition 3.5 in \cite{connes3}.

Connes makes the identifications of the $U(1)$ field $\Lambda$, $SU(2)$ field $Q$ and the $SU(3)$ field $V^{'}$ with gauge boson fields as follows,

\begin{equation}
\Lambda = \frac{ig_{1}}{2}\gamma^{\mu}B_{\mu}
\end{equation}

\begin{equation}
Q = \frac{ig_{2}}{2}\gamma^{\mu}W^{a}_{\mu}\sigma_{a}
\end{equation}

\begin{equation}
V^{'} = -\frac{ig_{3}}{g}A^{a}_{\mu}T_{a}
\end{equation}

where $B_{\mu}$ is the weak hypercharge field, $W_{\mu}$ is the W boson field and $A_{\mu}$ is the gluon field. $\sigma_{a}$ and $T_{a}$ are the Pauli matrices and Gell-Mann matrices respectively while $g_{1}$, $g_{2}$ and $g_{3}$ are the coupling constants for the weak hypercharge, W boson and gluon fields respectively, and are fixed at unification.

The Hilbert space $\mathcal{H}$ is given by the product $\mathcal{H} = L^{2}\left(M, S\right)\otimes \mathcal{H}_{F}$, where $\mathcal{H}_{F}$ is given a basis labelled by the elementary quarks,

$$
Q = \begin{pmatrix}
Q_{L} \\
 d_{R}\\
u_{R} 
\end{pmatrix}
$$

and the elementary leptons,

$$
\ell = \begin{pmatrix}
\ell_{L} \\
 e_{R}
\end{pmatrix}
$$

As per the proof of Proposition 3.9 in \cite{connes3} and the computation done in Section 3 of \cite{connes1}, the Dirac operators for the quarks and leptons are obtained as,

\begin{widetext}
$$
\begin{aligned}
&D_{q} = \\&\begin{bmatrix}
\gamma^{\mu}\left (D_{\mu} -\frac{ig_{2}}{2}W_{\mu} -\frac{ig_{1}}{6}B_{\mu} -\frac{ig_{3}}{2}A_{\mu}  \right ) & \gamma^{5}\otimes k^{d}H & \gamma^{5}\otimes i\sigma_{2}k^{u}H\\ 
\gamma^{5}\otimes \left ( k^{d} \right )^{*}H^{*} & \gamma^{\mu}\left (D_{\mu} +\frac{ig_{1}}{3}B_{\mu} -\frac{ig_{3}}{2}A_{\mu} \right ) & 0\\ 
\gamma^{5}\otimes i\sigma_{2}\left ( k^{u} \right )^{*}H^{*} & 0 & \gamma^{\mu}\left ( D_{\mu} + \frac{i2g_{1}}{3}B_{\mu}-\frac{ig_{3}}{2}A_{\mu}\right )
\end{bmatrix}\otimes\mathbb{I}_{3} \end{aligned}
$$
\end{widetext}

$$
D_{\ell} =\begin{bmatrix}
\gamma^{\mu}\left (D_{\mu} -\frac{ig_{2}}{2}W_{\mu} +\frac{ig_{1}}{2}B_{\mu} \right ) & \gamma^{5}\otimes k^{e}H \\
\gamma^{5}\otimes \left ( k^{e} \right )^{*}H^{*} &  \gamma^{\mu}\left ( D_{\mu} + ig_{1}B_{\mu}\right )
\end{bmatrix}\otimes\mathbb{I}_{3} 
$$

where the covariant gravitational Dirac operator $D_{\mu}$ is given by,

$$
D_{\mu} = \partial + \omega_{\mu}
$$

\subsection{The Spectral Action}
The formula for the Standard Model gravitational action is given by the spectral action. The spectral action is a sum of the spectral trace of the Dirac operator and the fermionic action, which is obtained by the operating of the Dirac operator on the Hilbert space basis. The spectral action is given by,

\begin{equation}
S = \mathrm{Tr}\left(f\left(\frac{D^{2}}{m^{2}_{0}}\right)\right) + \left(\psi, D\psi\right)
\end{equation}

where $f$ is a cutoff function and $m^{2}_{0}$ is a suitable unification scale, such that higher powers of the trace are suppressed by higer powers of $m^{2}_{0}$. The heat kernel trace of the above formula is given by the asymptotic expansion,

\begin{equation}
\mathrm{Tr}\left(f\left(P\right)\right) = \sum_{n}f_{n-4}a_{n}(P)
\end{equation}

where the $f_{n}$ are given by,

$$
f_{0} = \int_{0}^{\infty}f(u)u\mathrm{d}u
$$

$$
f_{2} = \int_{0}^{\infty}f(u)\mathrm{d}u
$$

$$
f_{2(n+2)} = (-1)^{n}f^{n}(0) \;\;\;\;\; n>\geq 0
$$

and the $a_{n}$ are known as the Seeley-DeWitt coefficients. In the above formula, due to the high value of $m^{2}_{0}$, only the first three coefficients need be considered, as the higher coefficients are suppressed by the higher powers of the energy scale. The bosonic part of the action may thus be given by,

\begin{equation}
\mathrm{Tr}\left(f\left(\frac{D^{2}}{m^{2}_{0}}\right)\right) = f_{0}m^{4}_{0}a_{0} + f_{2}m^{2}_{0}a_{2} + f_{4}a_{4} + 0\left(\frac{1}{m^{2}_{0}}\right)
\end{equation}

The odd coeffecients vanish for the case where the manifold has no boundary. For a general Laplacian operator of the form,

\begin{equation}
\Delta = -\nabla^{2} - E
\end{equation}

where $\nabla = \left(\partial + \omega^{'}\right)$ the first three coefficients are given by,

$$
a_{0} = \frac{1}{16\pi^{2}}\int \mathrm{vol}g 
$$

$$
a_{2} = \frac{1}{16\pi^{2}}\int E \;\;\mathrm{vol}g 
$$

$$
a_{4} = \frac{1}{192\pi^{2}}\int\left(6E^{2} + 2\nabla^{2}E + \omega^{'}_{\mu\nu}\omega^{'\mu\nu}\right) \mathrm{vol}g 
$$

where $\omega^{'}_{\mu\nu}$ is the curvature of $\omega^{'}$.

By computing the Heat kernel trace for the Dirac operator as has been done in the appendix of \cite{connes1}, the bosonic action, which is given by the heat kernel trace of the square of the Dirac operator $D$ upto an energy scale is obtained as,

\begin{widetext}
$$
\begin{aligned}
S_{B} = & \; \frac{45f_{4}m^{4}_{0}}{4\pi^{2}}\int \mathrm{d}^{4}x\sqrt{g} + \frac{3m^{2}_{0}}{4\pi^{2}}f_{2}\int\left(\frac{5}{6}R - 2yH^{*}H\right)\mathrm{d}^{4}x\sqrt{g}+\frac{5f_{4}}{640\pi^{2}}\int\left(12R^{\mu}_{;\mu} + 11R^{*}R^{*} - 18C_{\mu\nu\rho\lambda }C^{\mu\nu\rho\lambda }\right)\mathrm{d}^{4}x\sqrt{g}\\
&+\frac{3y^{2}f_{4}}{4\pi^{2}}\int\left(D_{\mu}H^{*}D^{\mu}H - \frac{1}{6}H^{*}H\right)\mathrm{d}^{4}x\sqrt{g}+\frac{f_{4}}{4\pi^{2}}\int\left(g^{2}_{3}A_{\mu\nu}A^{\mu\nu} + g^{2}_{2}W_{\mu\nu}W^{\mu\nu} + \frac{5}{3}g^{2}_{1}B_{\mu\nu}B^{\mu\nu}\right)\mathrm{d}^{4}x\sqrt{g}\\
&+\frac{f_{4}}{4\pi^{2}}\int\left(3z^{2}(H^{*}H)^{2} - y^{2}(H^{*}H)^{\mu}_{;\mu}\right)\mathrm{d}^{4}x\sqrt{g}
\end{aligned}
$$
\end{widetext}

where,

$$
y^{2} = \mathrm{Tr}\left(\left | k^{u} \right |^{2} + \left | k^{d} \right |^{2} + \frac{1}{3}\left | k^{e} \right |^{2}\right)
$$

$$
z^{2} = \mathrm{Tr}\left(\left(\left | k^{u} \right |^{2} + \left | k^{d} \right |^{2}\right)^{2} + \frac{1}{3}\left | k^{e} \right |^{4}\right)
$$

With the derivation of the above bosonic action, the formulation of non-commutative geometry as a mathematical artifice and it's application in particle physics is essentially complete. The use of non-commutative geometry in physics as a unifying link between gravity and the Standard Model is thus a purely geometric approach to achieving a true unified field theory, which must be thought of as an ingenious achievement by Alain Connes as far as the unification of physics is concerned. 

In the next section, we attempt to extend the noncommutative framework developed by Alain Connes into a new metric geometry, through the introduction of a new definition of the metric tensor, and in the process, attempt to develop gravity, gauge potentials, the Higgs boson and fermions as aspects of the curvature of spacetime.

\section{The Theory of Geometrodynamics}
\subsection{The generalised Derivative}
Before we proceed to define a new metric to build a foundation for a general geometry upon, we will consider a general type of derivative, particularly a covariant derivative and it's ramifications. Consider a general type of covariant derivative of the form,

\begin{equation}
\mathcal{D} = \partial + \frac{1}{2}\omega_{\mu}\mathrm{d}x^{\mu}\otimes \mathbb{I}_{N} + \frac{ig}{2}A_{\mu}\mathrm{d}x^{\mu}\otimes\mathbb{I}_{D} + \frac{1}{\alpha}\Phi\otimes\chi
\end{equation}

where $A_{\mu}$ is a gauge field obtained from the information afforded to us by a spectral triple, the $\mathrm{d}x^{\mu}$ are dimensionless unit vectors, $g$ is a coupling constant, $\chi$ is matrix valued and $\alpha$ is a constant. $\Phi$ is given by,

$$
\Phi = \begin{pmatrix}
c\sqrt{-1} & H\\ 
H^{*} & c\sqrt{-1}
\end{pmatrix}
$$

where $H$ is a Higgs field and $c$ is a constant. By writing the above covariant derivative in the form,

$$
\mathcal{D} = \partial + \mathcal{A}
$$

we define it's curvature as,

\begin{equation}
\mathcal{F} = \mathrm{d}\mathcal{A} + \mathcal{A}\wedge\mathcal{A}
\end{equation}

For the above covariant derivative, we obtain the curvature as,

\begin{widetext}
$$
\begin{aligned} 
\mathcal{F} = &\;\frac{1}{2}\mathbf{R}_{\mu\nu}\mathrm{d}x^{\mu}\wedge\mathrm{d}x^{\mu}\otimes \mathbb{I}_{N} + \frac{ig}{2}A_{\mu\nu}\mathrm{d}x^{\mu}\wedge\mathrm{d}x^{\mu}\otimes \mathbb{I}_{D} +   \frac{1}{\alpha}\begin{pmatrix}
0 & D_{\mu}H\\ 
D_{\mu}H^{*} & 0
\end{pmatrix}\mathrm{d}x^{\mu}\wedge\chi \\
&+  \frac{1}{\alpha^{2}}\begin{pmatrix}
HH^{*} - c^{2} & 0\\ 
0 & HH^{*} - c^{2}
\end{pmatrix}\chi\wedge\chi
\end{aligned}
$$
\end{widetext}

where $\mathbf{R}$ is the Riemann curvature tensor.

The squared curvature is then obtained by,

\begin{widetext}
\begin{equation}
\mathcal{F}^{2} = \frac{N}{4}\mathbf{R}^{\mu\nu}\mathbb{R}^{\mu\nu} - \frac{Dg^{2}}{4}A_{\mu\nu}A^{\mu\nu} + \frac{\eta}{\alpha^{2}}D_{\mu}H D^{\mu}H^{*}+ \frac{\eta^{2}}{\alpha^{4}}\left(HH^{*} - c^{2}\right)^{2} 
\end{equation}
\end{widetext}

where $\eta$ is a parameter defined by,

$$
\eta = \left \langle \chi,\chi \right \rangle
$$

Having worked out the ramifications of a general covariant derivative whose form is given in (4.23), we are now in a position to define a new form of a derivative, known as a generalised derivative. The generalised derivative is constructed out of the data obtained through the definition of a spectral triple. 

The spin connection is defined in a spectral triple through the heuristic definition of the Riemannian metric as has been given in equation (3.2). The gauge fields are parametrised by the continuous inner fluctuations $A^{1,0}$ obtained as the continuous part of the inner fluctuations of the metric. The discrete part $A^{0,1}$ of the inner fluctuations is parametrised by quaternion valued functions $H$ and $H^{*}$ which can be though of as the Higgs field. We thus construct the generalised derivative through the formula,

\begin{widetext}
\begin{equation}
\mathcal{D} = \partial + \frac{1}{2}\omega_{\mu}\mathrm{d}x^{\mu}\otimes\mathbb{I}_{N} - \sqrt{-1}A^{1,0}_{\mu}\mathrm{d}x^{\mu} \otimes\mathbb{I}_{D} + \frac{1}{\alpha}\begin{pmatrix}
c\sqrt{-1} & H\\ 
H^{*} & c\sqrt{-1}
\end{pmatrix}\otimes\chi
\end{equation}
\end{widetext}

where $\alpha$ and $c$ are constants while $\chi$ is matrix valued.

By parametrising $A^{1,0}$ such that it physically represents a gauge field, we are able to construct a yang-Mills Lagrangian by taking it's exterior covariant derivative. We do this by setting,

$$
A^{1,0}_{\mu} = \frac{g}{2}A_{\mu}
$$

where $A_{\mu}$ is a gauge boson field.

By using the above parametrisation and finding the square of the gauge curvature, we obtain the Lagrangian as,

\begin{widetext}
\begin{equation}
\mathcal{L} = \frac{N}{4}\mathbf{R}_{\mu\nu}\mathbf{R}^{\mu\nu} - \frac{Dg^{2}}{4}A_{\mu\nu}A^{\mu\nu} + \frac{\eta}{\alpha^{2}}D_{\mu}H D^{\mu}H^{*}+ \frac{\eta^{2}}{\alpha^{4}}\left(HH^{*} - c^{2}\right)^{2}
\end{equation}
\end{widetext}

Using constants $N_{R}$, $N_{D}$ and $N_{H}$, we reparametrise the above Lagrangian to obtain,

\begin{widetext}
\begin{equation}
\mathcal{L} = \frac{N}{4N^{2}_{R}}\mathbf{R}_{\mu\nu}\mathbf{R}^{\mu\nu} - \frac{Dg^{2}}{4N^{2}_{D}}A_{\mu\nu}A^{\mu\nu} + \frac{\eta}{\alpha^{2}N^{2}_{H}}D_{\mu}H D^{\mu}H^{*}- \frac{\eta^{2}}{\alpha^{4}N^{4}_{H}}\left(HH^{*} - c^{2}\right)^{2} + \mathrm{const}.
\end{equation}
\end{widetext}

The generalised derivative, purely constructed from the data afforded to us by the definition of a spectral triple, gives a curvature which completely determines the dynamics of gravitation, gauge interactions and the Higgs field, thus giving us a unified bosonic action. 

\subsection{The Generalised Space}
In order to develop and describe a new metric geometry, we must define a few terminologies and primitive notions. The most important of these is what we will call a generalised space, which is described by an ordered pair $\left(\mathcal{C}, \gamma\right)$, where $\mathcal{C}$ is a spectral triple and $\gamma$ is a generalised metric, which is not to be confused with the $\mathbb{Z}$ modulo 2 grading we described earlier. 

The properties and ramifications of the spectral triple have been described earlier. So, we will now describe the properties of the generalised metric, which as we shall soon see, will play a role in our new geometry which is quite analogous to that played by the Riemannian metric in Riemannian geometry.

To develop the idea of the generalised metric, we will first look at the relationship between the metric in Riemannian geometry and the veilbein. The veilbein $e^{a}_{\mu}$ may be thought of as the square root of the metric tensor and is given by the general formula,

\begin{equation}
g_{\mu\nu} = e^{a}_{\mu}e^{b}_{\nu}\eta_{ab}
\end{equation}

Using the veilbein, we may transform tensors to act along curved space or flat space co-ordinates. This allows an easy transformation between the two otherwise distinct physical systems.

The spin connection $\omega^{ab}_{\mu}$ is defined by the relations,

\begin{equation}
\omega^{ab}_{\mu} = e^{a}_{\nu}\partial_{\mu}e^{\nu a} + e^{a}_{\nu}e^{\lambda b}\Gamma^{\nu}_{\lambda\mu}
\end{equation}

\begin{equation}
\omega_{\mu} = \sigma_{ab}\omega^{ab}_{\mu}
\end{equation}

\begin{equation}
\sigma_{ab} = \frac{i}{2}\left [ \gamma_{a},\gamma_{b} \right ]
\end{equation}

As a compatibility condition, we require that the Riemannian covariant derivative, described using the Christoffel symbols, must vanish. This gives us the compatibility condition as,

\begin{equation}
\nabla_{\nu}e^{a}_{\mu} = 0
\end{equation}

\begin{equation}
\partial_{\nu}e^{a}_{\mu} - \eta_{bc}\omega^{ab}_{\nu}e^{c}_{\mu} + e^{\rho}_{d}\left(\partial_{\nu}e^{a}_{\mu} - \eta_{bc}\omega^{ab}_{\nu}e^{c}_{\mu}\right) e^{d}_{\rho} = 0
\end{equation}

Since $e^{\rho}_{d}e^{d}_{\rho} = N$, we have,

\begin{equation}
\left(1-N^{2}\right)\left(\partial_{\nu}e^{a}_{\mu} - \eta_{bc}\omega^{ab}_{\nu}e^{c}_{\mu}\right) = 0
\end{equation}

Contracting with $\sigma_{ab}$, we have,

\begin{equation}
\sigma_{ab}\partial_{\nu}e^{a}_{\mu} - \eta_{bc}\omega_{\nu}e^{c}_{\mu} = 0
\end{equation}

By making the substitution $\left \{ e^{a}_{\mu}, \omega_{\mu} \right \} \rightarrow \left \{ E^{a}_{\mu}, \mathcal{A}_{\mu} \right \}$ we have,

\begin{equation}
\sigma_{ab}\partial_{\nu}E^{a}_{\mu} - \eta_{bc}\mathcal{A}_{\nu}e^{c}_{\mu} = 0
\end{equation}

Solving for $E^{a}_{\mu}$, the veilbein of $\mathcal{A}_{\mu}$, we obtain a means to transform between a flat space co-ordinate system and a co-ordinate system wherein the spin connection, ie, the curvature connection of the manifold is $\mathcal{A}_{\mu}$.  we define the generalised metric $\gamma_{\mu\nu}$ as,

\begin{equation}
\gamma_{\mu\nu} = E^{a}_{\mu}E^{a}_{\nu}\eta_{ab}
\end{equation}

This new definition of a metric completely geometrises the bosonic mechanics described in the previous sections. The generalised Riemann curvature constructed out of this metric will now contain the Riemann curvature tensor of general relativity, the Yang-Mills field strength tensor and the Higgs boson dynamics. 

The new generalised metric is very important inasmuch as the nature covariance and contravariance is concerned. While in standard Riemannian geometry, indices are lowered and raised using the metric tensor, in the new geometry, this is done using the generalised metric. The inverse of the generalised metric $\gamma^{\mu\nu}$ is used to raise indices while the covariant generalised metric $\gamma_{\mu\nu}$ is used to lower indices. 

The new metric tensor will also make imperative the amending of several other laws and relations as far a geometry is concerned. For example, consider the relationship between the gamma matrices $\gamma_{\mu}$ and the metric tensor $g_{\mu\nu}$,

$$
\left \{ \gamma^{\mu}, \gamma^{\nu} \right \} = g_{\mu\nu}
$$

In place of this law, we must now define new Dirac matrices $\Gamma^{\mu}$ and introduce an analogous relationship between them and the generalised metric $\gamma_{\mu\nu}$. This relationship will be given by,

\begin{equation}
\left \{ \Gamma^{\mu}, \Gamma^{\nu} \right \} = \gamma_{\mu\nu}
\end{equation}

In Riemannian geometry, the volume of a given region of space is obtained by the integration of the volume form, i.e, the root of the negative of the metric determinant $g$ times the infinitesimal volume. This is given by,

$$
\mathrm{volume} = \int _{M}\sqrt{-g}\mathrm{d}^{4}x
$$

This definition of the volume of the manifold is replaced by a new definition using the square root of the negative of the determinant $\gamma$ of the generalised metric. This new definition is given by,

\begin{equation}
\mathrm{volume} = \int_{M}\sqrt{-\gamma}\mathrm{d}^{4}x
\end{equation}

\subsection{Generalised Spaces as Generalised Riemannian Spaces}
The new definition of the metric tensor allows us to re-evaluate the geometrical system offered to us by the mathematical apparatus of general relativity. We are now in a suitable position to expand upon the traditional Riemannian geometry of general relativity into the aforementioned generalised metric geometry. Since the dynamics of gauge fields, the Higgs field and gravity are contained in the Riemann curvature of the generalised metric, we can interpret the curvature of a space endowed with the generalised metric as these fields. Having made this observations, we will proceed onto our proposed generalisation of Riemannian geometry.

The cornerstone of Riemannian geometry is the line element, which is used to define the notion of distance using the metric tensor, which acts as a suitable generalisation of the flat space Euclidean distance to curved spaces. This serves as an adequate description of the infinitesimal distance for spaces with only gravitational curvature, but for our new generalisation, we replace the Riemannian distance formula with,

\begin{equation}
\mathrm{d}s^{2} = \gamma_{\mu\nu}\mathrm{d}x^{\mu}\mathrm{d}x^{\nu}
\end{equation}

The above formula allows us to recover the ordinary Riemannian distance formula if we make some specific simplyfying assumptions. Under the limit that $\mathcal{A}_{\mu}$ reduces to $\omega_{\mu}$, that is, the space is purely of a gravitational nature, then, the generalised metric $\gamma_{\mu\nu}$ obviously reduces to the Riemannian metric $g_{\mu\nu}$. Under this assumption, the above formula reduces to it's Riemannian counterpart.

The equations of a material particle moving along geodesics of this newly constructed space can be easily obtained by taking the extremum of the differential distance. To do this, we set,

$$
\delta\int\sqrt{ \gamma_{\mu\nu}\mathrm{d}x^{\mu}\mathrm{d}x^{\nu}} = 0
$$

By carrying out a standard calculation, we derive a generalised geodesic equation. The generalised geodesic equation is first order in the derivatives of the generalised metric and gives to us an expression which governs how particles move under the influence of the various fields on the manifold. The generalised geodesic equation takes the form,

\begin{equation}
\frac{\mathrm{d}^{2}x^{\mu}}{\mathrm{d}\tau^{2}} - \left \{ \mu \atop \alpha\beta \right \}\frac{\mathrm{d}x^{\mu}}{\mathrm{d}\tau}\frac{\mathrm{d}x^{\mu}}{\mathrm{d}\tau} = 0
\end{equation}

where the generalised Christoffel symbols of the second kind are given by a formula analogous to that of the Christoffel symbols of the second kind in Riemannian geometry, written as,

\begin{equation}
\left \{ \mu \atop \alpha\beta \right \} = \frac{1}{2}\gamma^{\lambda\mu}\left(\frac{\partial \gamma_{\lambda\alpha}}{\partial x^{\beta}} + \frac{\partial \gamma_{\lambda\beta}}{\partial x^{\alpha}} - \frac{\partial \gamma_{\alpha\beta}}{\partial x^{\lambda}}\right)
\end{equation}

Since in the Riemannian approximation, the generalised metric reduces to the Riemannian metric, the generalised Christoffel symbols reduce to the Riemann-Christoffel symbols at this limit as well. Hence, when the spin connection of the space co-incides with the Riemannian spin connection, the generalised geodesic equation reduces to the gravitational geodesic equation.

The generalised Riemann curvature is obtained by using an analogue of the formula used in Riemannian geometry. In terms of the generalised Christoffel symbols, the generalised Riemann curvature is obtained as,

\begin{widetext}
\begin{equation}
\widehat{R}^{\rho}_{\mu\nu\lambda} = \frac{\partial }{\partial x^{\nu}} \left \{ \rho \atop \lambda\mu \right \} - \frac{\partial }{\partial x^{\lambda}} \left \{ \rho \atop \nu\mu\right \}+ \left \{ \rho \atop \nu\sigma \right \}\left \{ \sigma \atop \lambda\mu \right \} - \left \{ \rho \atop \lambda\sigma \right \}\left \{ \sigma \atop \nu\mu \right \}
\end{equation}
\end{widetext}

The Ricci curvature and Ricci scalar are obtained by suitable contractions of the Riemann curvature. It is also necessary to note a specific choice of the co-ordinate system. Since the $\gamma_{\mu\nu}$ are purely geometric quantities, the components of the generalised metric are dependent solely on the choice of the co-ordinate system. As a result, it is possible to choose a system of co-ordinates which give the following results,

$$
\sqrt{-\gamma} = 1
$$

$$
\left \{ \beta\atop \beta\alpha\right \}  = 0
$$

Under these conditions, the Ricci curvature is obtained as a simple expression,

\begin{equation}
\widehat{R}_{\mu\nu} = -\frac{\partial}{\partial x^{\alpha}}\left \{ \alpha \atop \mu\nu\right \} + \left \{ \beta\atop \mu\alpha\right \}\left \{ \alpha \atop \nu\beta\right \}
\end{equation}

\subsection{A Universal Action and Hamilton's Principle}
The spectral action developed by Alain Connes encapsulates the dynamics of all fundamental interactions and fermions. Having developed a new metric geometry which encodes the same information in a purely geometric framework and allows for an interpretation of such phenomena as a manifestation of the curvature of the chosen manifold, we can proceed to geometrise the spectral action.

As defined earlier, we have a general operator on the manifold, which encodes the properties of gravity, gauge ptentials and the Higgs field, which are all obtained from the data contained in the definition of a spectral triple. This general operator, when squared, is given by,

\begin{equation}
\mathcal{D}^{2} = \left(\partial + \mathcal{A}\right)^{2}
\end{equation}

where $\mathcal{A}$, the collection of all the aforementioned fields is not to be confused with the algebra of co-ordinates that is contained in the spectral triple. As mentioned earlier, for a general operator $P$ of the form,

\begin{equation}
P = \mathcal{D}^{2} + E
\end{equation}

We can define the heat kernel functional, for a suitable function $f$ and a suitably high cut-off value $\Lambda^{2}_{E}$, as a sum of three terms, with the successive terms being effectively cancelled out by higher powers of the energy scale. This is given by,

\begin{equation}
\mathrm{Tr}\left(f\left(\frac{P}{\Lambda^{2}_{E}}\right)\right) = f_{4}\Lambda^{4}_{E}a_{0} + f_{2}\Lambda^{2}_{E}a_{2} + f_{0}a_{4} + \mathcal{O}\left(\frac{1}{\Lambda^{2}_{E}}\right)
\end{equation}

where the co-efficients $f_{n}$ are as given in Section 2 and the heat kernel co-efficients $a_{0}$, $a_{2}$ and $a_{4}$ are given by,

$$
a_{0} = \frac{1}{16\pi^{2}}\int\mathrm{vol}g
$$

$$
a_{2} = \frac{1}{16\pi^{2}}\int E\;\mathrm{vol}g
$$

$$
a_{4} = \frac{1}{192\pi^{2}}\int \left(6E^{2} + 2\mathcal{D}^{2}E + \mathcal{A}_{\mu\nu}\mathcal{A}^{\mu\nu}\right)\mathrm{vol}g.
$$

In our theory, $E$ is chosen to be zero and thus, the operator $P$ simply reduces to the square of the generalised derivative. The heat kernel expansion for this then becomes,

\begin{widetext}
$$
\begin{aligned}
\mathrm{Tr}\left(f\left(\frac{\mathcal{D}^{2}}{\Lambda^{2}_{E}}\right)\right) = &\; \frac{f_{4}\Lambda^{4}_{E}}{16\pi^{2}}\int\sqrt{-\gamma}\mathrm{d}^{4}x + \frac{f_{0}}{192\pi^{2}}\int  \left (\frac{N}{4N^{2}_{R}}\mathbf{R}_{\mu\nu}\mathbf{R}^{\mu\nu}  \right )\sqrt{-\gamma}\mathrm{d}^{4}x \\
& + \frac{f_{0}}{192\pi^{2}}\int  \left (- \frac{Dg^{2}}{4N^{2}_{D}}A_{\mu\nu}A^{\mu\nu} + \frac{\eta}{\alpha^{2}N^{2}_{H}}D_{\mu}H D^{\mu}H^{*}  \right )\sqrt{-\gamma}\mathrm{d}^{4}x \\
& + \frac{f_{0}}{192\pi^{2}}\int  \left (\frac{\eta^{2}}{\alpha^{4}N^{4}_{H}}\left(HH^{*} - c^{2}\right)^{2} + \lambda_{0} \right )\sqrt{-\gamma}\mathrm{d}^{4}x
\end{aligned}
$$
\end{widetext}

where $\lambda_{0}$ is the constant obtained as a result of the reparametrisation of the Lagrangian as done in (3.6).

The above action can be written more succintly as,

$$
\mathrm{Tr}\left(f\left(\frac{\mathcal{D}^{2}}{\Lambda^{2}_{E}}\right)\right) =  \int  \left (\tau_{0} + \frac{1}{2\kappa_{0}} \widehat{R}_{\mu\nu\rho\lambda}\widehat{R}^{\mu\nu\rho\lambda}   \right )\sqrt{-\gamma}\mathrm{d}^{4}x 
$$

where we have set,

$$
\tau_{0} = \frac{f_{4}\Lambda^{4}_{E}}{16\pi^{2}}
$$

$$
\frac{1}{2\kappa_{0}} = \frac{\sigma^{2}f_{0}}{192\pi^{2}}
$$

$$
\sigma^{2} = \sigma^{ab}\sigma_{ab}
$$
Thus, we have constructed the bosonic action purely out of the definition of the heat kernel, the geometrical apparatus and the information contained in the spectral triple. The complete action is now given by,

\begin{equation}
S = \mathrm{Tr}\left(f\left(\frac{\mathcal{D}^{2}}{\Lambda^{2}_{E}}\right)\right) + \left(\overline{\Psi}, D\Psi\right)
\end{equation}

Using Hamilton's principle, we can obtain a general field equation describing our system. According to Hamilton's principle,

\begin{equation}
\frac{\delta S}{\delta \gamma^{\mu\nu}} = 0
\end{equation}

This gives us a set of ten partial-differential field equations to govern our system. Upon carrying out the variation, we obtain the general field equations,

\begin{widetext}
\begin{equation}
\frac{\delta}{\delta \gamma^{\mu}}\left \{\frac{1}{2\kappa_{0}} \widehat{R}_{\mu\nu\rho\lambda}\widehat{R}^{\mu\nu\rho\lambda} \sqrt{-\gamma} + \tau_{0}\sqrt{-\gamma} \right \} = -\frac{\delta}{\delta \gamma^{\mu}}\left \{ \overline{\Psi}D\Psi \sqrt{-\gamma} \right \}
\end{equation}
\end{widetext}

We obtain the variations as,

$$
\frac{\delta}{\delta \gamma^{\mu\nu}}\left \{ \widehat{R}_{\mu\nu\rho\lambda}\widehat{R}^{\mu\nu\rho\lambda} \sqrt{-\gamma} \right \} = 4\widehat{R}_{\mu}^{\sigma\rho\lambda}\widehat{R}_{\nu\sigma\rho\lambda} + \frac{1}{2}\gamma_{\mu\nu}\widehat{R}_{\mu\nu\rho\lambda}\widehat{R}^{\mu\nu\rho\lambda} 
$$

$$
-2\frac{\delta}{\delta \gamma^{\mu\nu}}\left \{ \overline{\Psi}D\Psi \sqrt{-\gamma} \right \} = T_{\mu\nu}
$$

$$
\frac{\delta}{\delta \gamma^{\mu\nu}}\left \{ \tau_{0}\sqrt{-\gamma} \right \} = \frac{1}{2}\gamma_{\mu\nu}\tau_{0}.
$$

These variations when substituted into the total variation give us the field equations as,

\begin{equation}
4\widehat{R}_{\mu}^{\sigma\rho\lambda}\widehat{R}_{\nu\sigma\rho\lambda} + \frac{1}{2}\gamma_{\mu\nu}\widehat{R}_{\mu\nu\rho\lambda}\widehat{R}^{\mu\nu\rho\lambda} = \kappa_{0}T_{\mu\nu} - \gamma_{\mu\nu}\kappa_{0}\tau_{0}.
\end{equation}
\section{The Physical Phenomena}

\subsection{The Action for Gravity and The Standard Model}
Having defined the mathematics and the broad gemoetrical apparatus upon which we hope to derive the formulation of general relativity and the Standard Model, we will proceed to do so by using the data we have obtained earlier through the theory of spectral triples.

The generalised derivative for the gravitational Standard Model is given by,

\begin{widetext}
\begin{equation}
\mathcal{D} = \partial + \frac{1}{2}\omega_{\mu}\mathrm{d}x^{\mu}\otimes\mathbb{I}_{N} - \sqrt{-1}A^{1,0}_{\mu}\mathrm{d}x^{\mu} \otimes\mathbb{I}_{D} + \frac{1}{\alpha}\begin{pmatrix}
c\sqrt{-1} & H\\ 
H^{*} & c\sqrt{-1}
\end{pmatrix}\otimes\chi
\end{equation}
\end{widetext}

here, $H$ is a quaternion valued function, which we identify as the Higgs field and parametrises the discrete internal fluctuations. The continuous part $A^{1,0}$ of the inner fluctuations is parametrised by a $U(1)$ field $\Lambda$, an $SU(2)$ field $Q$ and a $U(3)$ field $V$. 

The curvature of this derivative will be given by,

\begin{widetext}
$$
\begin{aligned} 
\mathcal{F} = &\;\frac{1}{2}\mathbf{R}_{\mu\nu}\mathrm{d}x^{\mu}\wedge\mathrm{d}x^{\mu}\otimes \mathbb{I}_{N} - \sqrt{-1}A^{1,0}_{\mu\nu}\mathrm{d}x^{\mu}\wedge\mathrm{d}x^{\mu}\otimes \mathbb{I}_{D} +   \frac{1}{\alpha}\begin{pmatrix}
0 & D_{\mu}H\\ 
D_{\mu}H^{*} & 0
\end{pmatrix}\mathrm{d}x^{\mu}\wedge\chi \\
&+  \frac{1}{\alpha^{2}}\begin{pmatrix}
HH^{*} - c^{2} & 0\\ 
0 & HH^{*} - c^{2}
\end{pmatrix}\chi\wedge\chi
\end{aligned}
$$
\end{widetext}

First, we will compute the curvature of the purely gauge bosonic part of the covariant derivative. The gauge potential as given earlier by the definition will be given by,

\begin{equation}
A^{1,0}_{\mu\nu} = \Lambda_{\mu\nu} + Q_{\mu\nu}\otimes\mathbb{I}_{2} + V_{\mu\nu}\otimes\mathbb{I}_{3}
\end{equation}

For the $U(3)$ field however, on account of the earlier relation,

$$
V = -V' - \frac{1}{3}\Lambda \mathbb{I}_{3}
$$

we obtain a relation between the $U(3)$ gauge curvature and that of the $SU(3)$ and $U(1)$ fields,

\begin{equation}
V_{\mu\nu} = -V'_{\mu\nu} -\frac{1}{3}\Lambda_{\mu\nu} \mathbb{I}_{3} 
\end{equation}

By calculating the square of the above curvature, we obtain,

$$
V_{\mu\nu}V^{\mu\nu} = V'_{\mu\nu}V'^{\mu\nu} + \frac{1}{3}\Lambda_{\mu\nu}\Lambda^{\mu\nu}
$$

By using the following relations,

\begin{equation}
V^{'}_{\mu} = \frac{g_{3}}{2}A_{\mu}
\end{equation}

\begin{equation}
\Lambda_{\mu} = \frac{g_{1}}{2}B_{\mu}
\end{equation}

The relation becomes the sum of the gauge curvatures of the gluon field and the weak hypercharge field, with certain coefficients, as given by,

\begin{equation}
V_{\mu\nu}V^{\mu\nu} = -\frac{g^{2}_{3}}{4}A_{\mu\nu}A^{\mu\nu} - \frac{g^{2}_{1}}{12}B_{\mu\nu}B^{\mu\nu}
\end{equation} 

The $SU(2)$ part $Q$ of the inner fluctuations is also used to define the W-boson field as,

\begin{equation}
Q_{\mu} = \frac{g_{2}}{2}W_{\mu}
\end{equation}

giving us the relation,

\begin{equation}
Q_{\mu\nu}Q^{\mu\nu} = \frac{g^{2}_{2}}{4}W_{\mu\nu}W^{\mu\nu}
\end{equation}

Squaring the gauge boson curvature gives us the Lagrangian of the gauge bosons as,

\begin{equation}
A^{1,0}_{\mu\nu}A^{1,0 \mu\nu}=  \frac{3}{4}g^{2}_{1}B_{\mu\nu}B^{\mu\nu} + \frac{1}{4}g^{2}_{2}W_{\mu\nu}W^{\mu\nu} + \frac{3}{4}g^{3}_{2}G_{\mu\nu}G^{\mu\nu}
\end{equation}

The derivative of the Higgs field will be given by,

\begin{equation}
D_{\mu}H = \left(\partial_{\mu} - \frac{ig_{1}}{2}B_{\mu} - \frac{ig_{2}}{2}W_{\mu}\right)H
\end{equation}

The square of the curvature, reparamatrised with constants $N_{R}$, $N_{B}$, $N_{W}$, $N_{G}$ and $N_{H}$ is obtained as,

\begin{widetext}
$$
\begin{aligned}
\mathcal{F}^{2} = &\;\frac{1}{N^{2}_{R}}\mathbf{R}_{\mu\nu}\mathbf{R}^{\mu\nu}- \frac{3}{4N^{2}_{B}}g^{2}_{1}B_{\mu\nu}B^{\mu\nu} - \frac{1}{4N^{2}_{W}}g^{2}_{2}W_{\mu\nu}W^{\mu\nu} - \frac{3}{4N^{2}_{G}}g^{3}_{2}G_{\mu\nu}G^{\mu\nu} \\
&+ \frac{\eta}{\alpha^{2}N^{2}_{H}}D_{\mu}H D^{\mu}H^{*}- \frac{\eta^{2}}{\alpha^{4}N^{4}_{H}}\left(HH^{*} - c^{2}\right)^{2} + \lambda_{0}
\end{aligned}
    $$
\end{widetext}

The bosonic action, upon evaluation of the heat kernel is obtained as,

\begin{widetext}
$$
\begin{aligned}
S_{B} = &\;\frac{12f_{4}\Lambda^{4}_{E} + f_{0}\lambda_{0}}{192\pi^{2}}\int\sqrt{-\gamma}\mathrm{d}^{4}x\\
&+\frac{f_{0}}{192\pi^{2}}\int\left(\frac{1}{N^{2}_{R}}\mathbf{R}_{\mu\nu}\mathbf{R}^{\mu\nu}- \frac{3}{4N^{2}_{B}}g^{2}_{1}B_{\mu\nu}B^{\mu\nu} - \frac{1}{4N^{2}_{W}}g^{2}_{2}W_{\mu\nu}W^{\mu\nu}\right)\sqrt{-\gamma}\mathrm{d}^{4}x\\
&+\frac{f_{0}}{192\pi^{2}}\int\left(- \frac{3}{4N^{2}_{G}}g^{3}_{2}G_{\mu\nu}G^{\mu\nu} + \frac{\eta}{\alpha^{2}N^{2}_{H}}D_{\mu}H D^{\mu}H^{*}\right)\sqrt{-\gamma}\mathrm{d}^{4}x\\
&-\frac{f_{0}}{192\pi^{2}}\int\left( \frac{\eta^{2}}{\alpha^{4}N^{4}_{H}}\left(HH^{*} - c^{2}\right)^{2}\right)\sqrt{-\gamma}\mathrm{d}^{4}x
\end{aligned}
$$
\end{widetext}

To simplify the above action, we make a few substitutions, which are,

$$
\frac{12f_{4}\Lambda^{4}_{E} + f_{0}\lambda_{0}}{192\pi^{2}} = \delta_{0}
$$

$$
\frac{f_{0}}{192\pi^{2}N^{2}_{R}} = \alpha_{0}
$$

$$
\frac{3f_{0}g^{2}_{1}}{N^{2}_{B}768\pi^{2}} = \frac{3f_{0}g^{2}_{3}}{N^{2}_{G}768\pi^{2}} = \frac{f_{0}g^{2}_{2}}{N^{2}_{W}768\pi^{2}} = \frac{1}{4}
$$

$$
\sqrt{\frac{\eta f_{0}}{\alpha^{2}N^{2}_{H}192\pi^{2}}}H = \mathbf{H}
$$

$$
\sqrt{\frac{\eta f_{0}}{\alpha^{2}N^{2}_{H}192\pi^{2}}}c = z
$$

$$
\frac{192\pi^{2}}{f_{0}} = \mu_{0}
$$

This transforms the bosonic action into,

\begin{widetext}
$$
\begin{aligned}
S_{B} = &\; \int\left(\alpha_{0}\mathbf{R}_{\mu\nu}\mathbf{R}^{\mu\nu}- \frac{1}{4}B_{\mu\nu}B^{\mu\nu} - \frac{1}{4}W_{\mu\nu}W^{\mu\nu}- \frac{1}{4}G_{\mu\nu}G^{\mu\nu}\right)\sqrt{-\gamma}\mathrm{d}^{4}x\\
&+\int\left( \left | D_{\mu}\mathbf{H} \right |^{2}-\mu_{0}\left(\left | \mathbf{H} \right |^{2} - z^{2}\right)^{2}+\delta_{0}\right)\sqrt{-\gamma}\mathrm{d}^{4}x
\end{aligned}
$$
\end{widetext}

It is important to keep in mind that higher order terms and the resulting non-trivial couplings can be neglected only on account of the magnitude of the energy scale. If the energy scale were below the required unification scale, these terms, that is, the heat kernel co-oefficients $a_{6}$ onwards cannot be neglected. This implies that at low unification scales, there are additional non-trivial couplings between the various bosonic fields.

Thus, we have managed to recover the complete Standard Model and Yang-Mills gravity Lagrangian purely from the information contained in our new metric geometry. 

\subsection{The Gravitational Standard Model Field Equations}
Having derived the complete bosonic geometric action, we can now proceed to find the field equations for the particular case of the Standard Model. The field equations that we derive will have a very important physical interpretation. Just as the geodesic equation describes how the particles move under the influence of the various bosonic fields, the field equations will give us how the bosonic fields are related to, or are generated as a result of the spacetime curvature caused by these material particles. 

The field equations are obtained by the application of Hamilton's principle. Thus, the field equations derived in this section will be of a purely classical nature, and will be incapable of describing how these fields behave under a quantum framework. By applying Hamilton's Principle, we obtain the following equation,

\begin{equation}
\frac{\delta}{\delta\gamma^{\mu\nu}}\left \{ \mathrm{Tr}\left ( f\left ( \frac{\mathcal{D}^{2}}{\Lambda^{2}_{E}}  \right ) \right )+ \left(\overline{\Psi},D\Psi \right )\right \}
\end{equation}

Upon taking the variation, we obtain the following field equation,

\begin{equation}
\frac{\delta \mathcal{L}_{B}\sqrt{-\gamma}}{\delta\gamma^{\mu\nu}} = \frac{1}{2}T_{\mu\nu}
\end{equation}

where we have set the energy-momentum tensor of matter $T_{\mu\nu}$ as,

$$
T_{\mu\nu} = -2\frac{\delta}{\delta\gamma^{\mu\nu}}\left \{ \overline{\Psi}D\Psi\sqrt{-\gamma}\right \}
$$

and the Lagrangian $\mathcal{L}_{B}$ is defined by,

$$
S_{B} = \int\mathcal{L}_{B}\sqrt{-\gamma}\mathrm{d}^{4}x
$$

To drive the field equation, we must now carry out the variation of the bosonic Lagrangian. The variation is obtained as,

$$
\frac{\delta \mathcal{L}_{B}\sqrt{-\gamma}}{\delta\gamma^{\mu\nu}} = \frac{\delta \mathcal{L}_{B}}{\delta\gamma^{\mu\nu}} - \frac{1}{2}\gamma_{\mu\nu}\mathcal{L}_{B}
$$

The variations of the Riemannian and gauge boson Lagrangians are obtained as,

$$
\frac{\delta}{\delta\gamma^{\mu\nu}}\left \{ \alpha_{0}\mathbf{R}_{\mu\nu}\mathbf{R}^{\mu\nu}\right \} = 2\alpha_{0}\mathbf{R}^{\alpha}_{\mu}\mathbf{R}_{\nu\alpha}
$$

$$
\frac{\delta}{\delta\gamma^{\mu\nu}}\left \{ -\frac{1}{4}B_{\mu\nu}B^{\mu\nu}\right \} = -\frac{1}{2}B^{\alpha}_{\mu}B_{\nu\alpha}
$$

$$
\frac{\delta}{\delta\gamma^{\mu\nu}}\left \{ -\frac{1}{4}W_{\mu\nu}W^{\mu\nu}\right \} = -\frac{1}{2}W^{\alpha}_{\mu}W_{\nu\alpha}
$$

$$
\frac{\delta}{\delta\gamma^{\mu\nu}}\left \{ -\frac{1}{4}G_{\mu\nu}G^{\mu\nu}\right \} = -\frac{1}{2}G^{\alpha}_{\mu}G_{\nu\alpha}
$$

Now, we will proceed with the variation of the Higgs field. Since the Higgs potential is a scalar, it's variation vanishes. As a result, we only have the variation of the Higgs kinetic term. Thus, we obtain,

$$
\frac{\delta}{\delta\gamma^{\mu\nu}}\left \{ \left | D_{\mu}\mathbf{H} \right |^{2}-\mu_{0}\left(\left | \mathbf{H} \right |^{2} - z^{2}\right)^{2}\right \} = D_{\mu}\mathbf{H}^{*}D_{\nu}\mathbf{H}
$$

The above variation is clearly not symmetric. By symmetrising it, we obtain the symmetric tensor as,

$$
 D_{\mu}\mathbf{H}^{*}D_{\nu}\mathbf{H} = \frac{1}{2} D_{\mu}\mathbf{H}^{*}D_{\nu}\mathbf{H} + \frac{1}{2} D_{\nu}\mathbf{H}^{*}D_{\mu}\mathbf{H}
$$

The variation of the cosmological constant $\delta_{0}$ vanishes on account of the fact that it is a constant value. The final field equation is then obtained as,

\begin{equation}
N_{\mu\nu} - \frac{1}{2}\gamma_{\mu\nu}\mathcal{L}_{B} = \frac{1}{2}T_{\mu\nu}
\end{equation}

where,

$$
\begin{aligned} 
N_{\mu\nu} = &\; 2\alpha_{0}\mathbf{R}^{\alpha}_{\mu}\mathbf{R}_{\nu\alpha} -\frac{1}{2}B^{\alpha}_{\mu}B_{\nu\alpha}-\frac{1}{2}W^{\alpha}_{\mu}W_{\nu\alpha}\\
& -\frac{1}{2}G^{\alpha}_{\mu}G_{\nu\alpha} + \frac{1}{2} D_{\mu}\mathbf{H}^{*}D_{\nu}\mathbf{H} + \frac{1}{2} D_{\nu}\mathbf{H}^{*}D_{\mu}\mathbf{H}.
\end{aligned}
$$

\subsection{General Relativity as a Riemannian Approximation}
As our theory is a geometric realisation of a generalisation of relativity, it is imperative that at the relevant limits, we obtain Connes' theory and can derive general relativity from that as the Riemannian limit. To do this, we must consider the ramifications of the generalised derivative and how it can be modified and under what conditions it can reproduce general relativity in the Riemannian limit. Consider the case where the spin connection of the manifold co-incides with the Riemannian spin connection. In this, case, the generalised metric reduces to the Riemannian metric and the geodesic equation is obtained as,

$$
\frac{\mathrm{d}^{2}x^{\mu}}{\mathrm{d}\tau^{2}} - \Gamma^{\mu}_{\alpha\beta}\frac{\mathrm{d}x^{\mu}}{\mathrm{d}\tau}\frac{\mathrm{d}x^{\mu}}{\mathrm{d}\tau} = 0
$$

The action will also reduce to a Riemannian action and will take the form,

\begin{widetext}
$$
\begin{aligned}
S = & \;\int\left(\alpha_{0}\mathbf{R}_{\mu\nu}\mathbf{R}^{\mu\nu}- \frac{1}{4}B_{\mu\nu}B^{\mu\nu} - \frac{1}{4}W_{\mu\nu}W^{\mu\nu}- \frac{1}{4}G_{\mu\nu}G^{\mu\nu}\right)\sqrt{-g}\mathrm{d}^{4}x\\
&+\int\left( \left | D_{\mu}\mathbf{H} \right |^{2}-\mu_{0}\left(\left | \mathbf{H} \right |^{2} - z^{2}\right)^{2}+\delta_{0}\right)\sqrt{-g}\mathrm{d}^{4}x\\
&+ \int\overline{\Psi}D\Psi\sqrt{-g}\mathrm{d}^{4}x
\end{aligned}
$$
\end{widetext}

The above action reproduces a Yang-Mills type Riemannian theory that describes the behaviour of gravity with an action that is quadratic in the Riemann tensor. To recover general relativity from our framework, we can perturb the generalised derivative to account for the covariant derivative as well. We can achieve this by setting,

$$
\mathcal{D}_{\Lambda} = \nabla + \mathcal{D} 
$$

Here, $\nabla$ is the covariant derivative on the manifold, in terms of the generalised Christoffel symbols. This gives us the heat kernel expansion as,

\begin{widetext}
$$
\begin{aligned}
S_{B} = &\;\frac{12f_{4}\Lambda^{4}_{E} + f_{0}\lambda_{0}}{192\pi^{2}}\int\sqrt{-g}\mathrm{d}^{4}x+ \frac{f_{2}\Lambda^{2}_{E}}{64\pi^{2}}\int\widehat{R} \sqrt{-g}\mathrm{d}^{4}x \\
&+\frac{f_{0}}{192\pi^{2}}\int\left(\frac{1}{N^{2}_{R}}\mathbf{R}_{\mu\nu}\mathbf{R}^{\mu\nu}- \frac{3}{4N^{2}_{B}}g^{2}_{1}B_{\mu\nu}B^{\mu\nu} - \frac{1}{4N^{2}_{W}}g^{2}_{2}W_{\mu\nu}W^{\mu\nu}\right)\sqrt{-g}\mathrm{d}^{4}x\\
&+\frac{f_{0}}{192\pi^{2}}\int\left(- \frac{3}{4N^{2}_{G}}g^{3}_{2}G_{\mu\nu}G^{\mu\nu} + \frac{\eta}{\alpha^{2}N^{2}_{H}}D_{\mu}H D^{\mu}H^{*}\right)\sqrt{-g}\mathrm{d}^{4}x-\frac{f_{0}}{192\pi^{2}}\int\left( \frac{\eta^{2}}{\alpha^{4}N^{4}_{H}}\left(HH^{*} - c^{2}\right)^{2}\right)\sqrt{-g}\mathrm{d}^{4}x\\
&+ f_{0}\int\left(\frac{1}{480\pi^{2}}\widehat{R}^{\mu}_{;\mu} + \frac{1}{1152\pi^{2}}\widehat{R}^{2} \right)\sqrt{-g}\mathrm{d}^{4}x-\frac{f_{0}}{2880\pi^{2}}\int\left(\widehat{R}^{\mu\nu}\widehat{R}_{\mu\nu} + \widehat{R}^{\mu\nu\rho\lambda}\widehat{R}_{\mu\nu\rho\lambda}\right)\sqrt{-g}\mathrm{d}^{4}x
\end{aligned}
$$
\end{widetext}

At the Riemannian limit, that is, when the spin connection of the manifold is the same as the Riemannian spin connection, the following limiting results hold true,

$$
\gamma_{\mu\nu} = g_{\mu\nu}
$$

$$
\widehat{R}_{\mu\nu\rho\lambda} = R_{\mu\nu\rho\lambda}
$$

$$
\widehat{R}_{\mu\nu} = R_{\mu\nu}
$$

$$
\widehat{R} = R
$$

The bosonic action now reduces to an Einstein-Hilbert action complemented with terms quadratic in the Riemann curvature, Ricci tensor and Ricci scalar and the Lagrangian of the Standard Model. What this tells us is that in the case where we perturb the generalised metric to account for the covariant derivative as well, we are able to reproduce a Riemannian action which describes the same dynamics that are described by the Einstein-Hilbert action along with the Standard Model. If we make the following replacements,

$$
\frac{f_{0}}{2880\pi^{2}} = \beta_{0}
$$

$$
\frac{f_{0}}{480\pi^{2}} = \eta_{0} 
$$

$$
\frac{f_{0}}{1152\pi^{2}} = \zeta_{0} 
$$

$$
\frac{12f_{4}\Lambda^{4}_{E} + f_{0}\lambda_{0}}{192\pi^{2}} = \delta_{0}
$$

$$
\frac{f_{0}}{192\pi^{2}N^{2}_{R}} = \alpha_{0}
$$

$$
\frac{3f_{0}g^{2}_{1}}{N^{2}_{B}768\pi^{2}} = \frac{3f_{0}g^{2}_{3}}{N^{2}_{G}768\pi^{2}} = \frac{f_{0}g^{2}_{2}}{N^{2}_{W}768\pi^{2}} = \frac{1}{4}
$$

$$
\sqrt{\frac{\eta f_{0}}{\alpha^{2}N^{2}_{H}192\pi^{2}}}H = \mathbf{H}
$$

$$
\sqrt{\frac{\eta f_{0}}{\alpha^{2}N^{2}_{H}192\pi^{2}}}c = z
$$

$$
\frac{192\pi^{2}}{f_{0}} = \mu_{0}
$$

The bosonic action takes the much simpler form of,

\begin{widetext}
$$
\begin{aligned}
S_{B} = &\; \frac{f_{2}\Lambda^{2}_{E}}{64\pi^{2}}\int R \sqrt{-g}\mathrm{d}^{4}x+\int\left(\alpha_{0}\mathbf{R}_{\mu\nu}\mathbf{R}^{\mu\nu}- \frac{1}{4}B_{\mu\nu}B^{\mu\nu} - \frac{1}{4}W_{\mu\nu}W^{\mu\nu}- \frac{1}{4}G_{\mu\nu}G^{\mu\nu}\right)\sqrt{-g}\mathrm{d}^{4}x\\
&+\int\left( \left | D_{\mu}\mathbf{H} \right |^{2}-\mu_{0}\left(\left | \mathbf{H} \right |^{2} - z^{2}\right)^{2}+\delta_{0}\right)\sqrt{-g}\mathrm{d}^{4}x\\
&+ \int\left(\eta_{0}R^{\mu}_{;\mu} + \zeta_{0}R^{2}- \beta_{0}R^{\mu\nu}R_{\mu\nu} - \beta_{0}R^{\mu\nu\rho\lambda}R_{\mu\nu\rho\lambda}\right)\sqrt{-g}\mathrm{d}^{4}x\\
\end{aligned}
$$
\end{widetext}

Thus, we see that we have recovered Alain Connes' theory by taking the Riemannian limit of our theory. Thus, it is clear that at the unification scale, when we consider the geometry to have a Riemannian nature, we will recover Alain Connes' theory at the limit. By neglecting Riemannian terms of order two and above, we can reduce even this to account only for the Einstein-Hilbert action and the Standard Model. 

In order to fully work out the ramifications of this limiting case, we must consider the field equations as well and find the co-efficient of the Ricci scalar in the action in order for the theory to correctly account for Einstein's theory and reduce to Newton's theory correctly at the weak-field limit. The field equations will be obtained by once again applying Hamilton's principle, and will come out as,

\begin{equation}
\frac{f_{2}\Lambda^{2}_{E}}{64\pi^{2}} \left(R_{\mu\nu} - \frac{1}{2}g_{\mu\nu}R\right) = \frac{1}{2}T_{\mu\nu} - N_{\mu\nu} + \frac{1}{2}g_{\mu\nu}\mathcal{L}_{B} + E_{\mu\nu}
\end{equation}

where we have,

$$
T_{\mu\nu} = -2\frac{\delta}{\delta\gamma^{\mu\nu}}\left \{ \overline{\Psi}D\Psi\sqrt{-g}\right \}
$$

\begin{widetext}
$$
\begin{aligned} 
E_{\mu\nu} = &\; -\frac{\delta}{\delta g^{\mu\nu}}\left \{\eta_{0}R^{\mu}_{;\mu} + \zeta_{0}R^{2}- \beta_{0}R^{\mu\nu}R_{\mu\nu} - \beta_{0}R^{\mu\nu\rho\lambda}R_{\mu\nu\rho\lambda}  \right \}\\
&+ \frac{1}{2}g_{\mu\nu}\left \{\eta_{0}R^{\mu}_{;\mu} + \zeta_{0}R^{2}- \beta_{0}R^{\mu\nu}R_{\mu\nu} - \beta_{0}R^{\mu\nu\rho\lambda}R_{\mu\nu\rho\lambda}  \right \}
\end{aligned}
$$
\end{widetext}

This field equation now describes how the Einstein tensor is related to the total energy-momentum of other fields. For the above field equation to be true, we must have that,

$$
\frac{f_{2}\Lambda^{2}_{E}}{64\pi^{2}} = \frac{c^{4}}{16\pi}
$$

This gives us the unification energy scale, in the case that we must have general relativity as the Riemannian approximation as,

$$
\Lambda^{2}_{E} = \frac{4\pi c^{4}}{f_{2}}
$$

With this final result, the development of the theory of geometrodynamics in non commutative spaces is, as a logical system complete. the development of a metric geometry to describe fundamental interactions renders the use of co-ordinates a purely mathematical artifice, which while implemented in case of the purely gravitational field in case of general relativity, can now be extended to the total field as well. This allows us to have the freedom to choose specific co-ordinate systems which will allow the vanishing of these fields, which allows for us to consider the dynamics of these fields, at this point, in this mathematical framework, purely geometric phenomena.

\section*{Acknowledgements}
This work was carried out when the author was visiting the Centre for Fundamental Research and Creative Education in Bangalore. The author also wishes to thank Dr. B S Ramachandra for useful suggestions and advice regarding the manuscript.


\begin{thebibliography}{}
\bibitem{connes1} 
Ali Chamseddine, Alain Connes
\textit{The Spectral action principle}. 
Commun.Math.Phys. 186 (1997) 731-750 hep-th/9606001.

\bibitem{connes2} 
Alain Connes
\textit{Gravity coupled with matter and foundation of noncommutative geometry}. 
Commun.Math.Phys. 182 (1996) 155-176 hep-th/9603053

\bibitem{gou1} 
Hanying Guo, Jianming Li, Ke Wu
\textit{Standard model with Higgs as gauge field on fourth homotopy group}. 
Commun.Theor.Phys. 29 (1998) 93-98 hep-th/9408152 AS-ITP-94-26

\bibitem{ein1} 
Albert Einstein
\textit{The Foundation of the General Theory of Relativity}. 
Annalen Phys. 49 (1916) 769-822,  Annalen Phys. 14 (2005) 517-571

\bibitem{vas1}
D.V. Vassilevich
  \textit{Heat kernel expansion: User's manual}
  Phys.Rept. 388 (2003) 279-360 hep-th/0306138
 
 \bibitem{nov1}
S.F. Novaes
  \textit{Standard model: An Introduction}
 In *Sao Paulo 1999, Particles and fields* 5-102 hep-ph/0001283 IFT-P-010-2000

\bibitem{ein2}
Albert Einstein
  \textit{The Field Equations of Gravitation}
Sitzungsber.Preuss.Akad.Wiss.Berlin (Math.Phys.) 1915 (1915) 844-847 8

\bibitem{connes3}
Ali Chamseddine, Alain Connes, Matilde Marcolli
  \textit{Gravity and the standard model with neutrino mixing}
Adv.Theor.Math.Phys. 11 (2007) 991-1089 hep-th/0610241 


\bibitem{connes4}
Ali Chamseddine, Alain Connes
  \textit{Universal formula for noncommutative geometry actions: Unification of gravity and the standard model}
Phys.Rev.Lett. 77 (1996) 4868-4871
\bibitem{vas2}
D.V. Vassilevich
  \textit{Non-commutative heat kernel}
Lett.Math.Phys. 67 (2004) 185-194 hep-th/0310144

\bibitem{ein3}
Einstein, Albert
  \textit{On the Electrodynamics of Moving Bodies}
Annalen Phys. 17 (1905) 891-921, Annalen Phys. 14 (2005) 194-224

\bibitem{brod}
D.C. Brody, L.P. Hughston
  \textit{Geometric Quantum Mechanics}
Journal of Geometry and Physics 38: 1953, quant-ph/9906086

\bibitem{oni1}
A.L.Onishchik
  \textit{Fubini-Study metric}
in Hazewinkel, Michiel,
Encyclopedia of Mathematics, Springer, ISBN 978-1-55608-010-4.

\end{thebibliography}
\end{document}